\documentclass[twocolumn,pra,showpacs]{revtex4}
\usepackage{amsmath}
\usepackage{amssymb}
\usepackage{graphicx}
\usepackage{bm}

\begin{document}

\title{Optical properties of topological insulator Bragg gratings: Faraday rotation enhancement for $TM$ polarized light at large incidence angles}

\author{J. A. Crosse}
\email{alexcrosse@gmail.com}
\affiliation{Department of Electrical and Computer Engineering, National University of Singapore, 4 Engineering Drive 3, Singapore 117583.}

\date{\today}

\begin{abstract}
Using the transfer matrix formalism, we study the transmission properties of a Bragg grating constructed from a layered axionic material. Such a material can be realized by a topological insulator subject to a time-symmetry breaking perturbation, such as an external magnetic field or surface magnetic impurities. Whilst the reflective properties of the structure are only negligibly changed by the presence of the axionic material, the grating induces a Faraday rotation and ellipticity in the transmitted light. We find that for $TM$ polarized light incident on a $16$ layer structure at $76^{\mathrm{o}}$ to the normal the Faraday rotation can approach $\approx 232\,\mathrm{mrad}$ $(\approx 13^{\mathrm{o}})$, whilst interference from the multi-layered structure ensures high transmission. This is significantly higher than Faraday rotations for the $TM$ polarization at normal incidences or the $TE$ polarization at any incident angle. Thus, Bragg gratings in this geometry show a strong optical signal of the magneto-electric and, hence, provide an ideal system in which to observe this effect by optical means.
\end{abstract}


\maketitle

\section{Introduction}

Topological insulators are materials that display non-trivial topological order. They are time-reversal symmetric and, though insulating in the bulk, support protected conducting edge states \cite{TIrev1,TIrev2}. Such phenomenology can be found in Group V and Group V/VI alloys that display strong enough spin-orbit coupling to induce band inversion, for example $\mathrm{Bi_{1-x}Sb_{x}}$ \cite{3DTI1,3DTI2}, $\mathrm{Bi_{2}Se_{3}}$, $\mathrm{Bi_{2}Te_{3}}$ and $\mathrm{Sb_{2}Te_{3}}$ \cite{3DTI3,3DTI4}.

Initially studied for their unusual electronic properties, topological insulators also display some interesting electromagnetic effects, the most significant of which is the magneto-electric effect which induces mixing between the electric, $\mathbf{E}$, and magnetic induction, $\mathbf{B}$, fields at the surface of the material \cite{monopole,TITFT}. This effect can be used to realize an axionic material \cite{obukhov,wilczek}. In order to do this one must introduce a time-symmetry breaking perturbation of sufficient size to open a gap in the surface states, thereby converting the material from a surface conductor to a full insulator \cite{TITFT, monopole}. This can be achieved by introducing magnetic dopants to the surface \cite{monopole} (Current experiments have achieved a $\approx 100$ meV gap with $22\%$ $\mathrm{Cr}$ dopants \cite{cr}) or by the application of an external static magentic field \cite{qi}. In such a situation, for wavelengths below the bandgap, the usual constitutive relations become \cite{TITFT,chang}
\begin{align}
\mathbf{D}(\mathbf{r},\omega) &= \varepsilon(\mathbf{r},\omega)\mathbf{E}(\mathbf{r},\omega)+ \frac{\alpha}{\pi}\Theta(\mathbf{r},\omega)\mathbf{B}(\mathbf{r},\omega),\label{con1} \\
\mathbf{H}(\mathbf{r},\omega) &= \frac{1}{\mu(\mathbf{r},\omega)}\mathbf{B}(\mathbf{r},\omega) - \frac{\alpha}{\pi}\Theta(\mathbf{r},\omega)\mathbf{E}(\mathbf{r},\omega),\label{con2} 
\end{align}
where $\alpha$ is the fine structure constant and $\varepsilon(\mathbf{r},\omega)$, $\mu(\mathbf{r},\omega)$ and $\Theta(\mathbf{r},\omega)$ are the dielectric permittivity, magnetic permeability and axion coupling respectively, the latter of which takes even multiples of $\pi$ in a conventional magneto-dielectric and odd multiples of $\pi$ in an axionic material, with the magnitude and sign of the multiple given by the strength and direction of the time symmetry breaking perturbation. Here and in the following we use natural units with $\varepsilon_{0} = \mu_{0} = c = 1$. Physically, the axionic term is generated by a quantized Hall effect on the surface of the insulator \cite{TITFT,monopole}. The lowest Hall plateau results in $\Theta(\mathbf{r},\omega) = \pm\pi$. The axion coupling will not increase until the magnet perturbation is strong enough to access the next Hall plateau. Thus, such axion couplings of are only achievable with magnetic perturbations on the order of Tesla. Lower fields lead to higher plateaus. However, these plateaus are narrower and very high plateaus can be difficult to resolve.

Using Maxwell's equations, which are unchanged in an axionic material, with the constitutive relations in Eqs. \eqref{con1} and \eqref{con2}, one can show that the frequency components of the electric field obey the inhomogeneous Helmholtz equation
\begin{multline}
\bm{\nabla}\times\frac{1}{\mu(\mathbf{r},\omega)}\bm{\nabla}\times\mathbf{E}(\mathbf{r},\omega)-\omega^2\varepsilon(\mathbf{r},\omega)\mathbf{E}(\mathbf{r},\omega)\\ 
-i\omega\frac{\alpha}{\pi}\left[\bm{\nabla}\Theta(\mathbf{r},\omega)\times\mathbf{E}(\mathbf{r},\omega)\right] = i\omega\mathbf{J}(\mathbf{r},\omega),
\label{EHelmholtz}
\end{multline}
where $\mathbf{J}(\mathbf{r},\omega)$ is the usual current source. If the axion coupling is homogeneous, $\Theta(\mathbf{r},\omega) = \Theta(\omega)$, then the last term on the left-hand side vanishes and one finds that the propagation of the electric field is the same as in a conventional magneto-dielectric. As a result, electromagnetic waves propagating within a homogeneous axionic material retain there usual properties - dispersion is linear, the phase and group velocities are proportional to the usual refractive index, the fields are transverse and orthogonal polarizations do not mix. Thus, the effects of the axion coupling are only felt when the axion coupling varies in space. For layered, homogeneous media this will only occur at the interfaces where the properties of the medium change. 

The change in the axion coupling at a topological insulator/magneto-dielectric interface is predicted to causes a Faraday (transmission) rotation of $\approx 1-10\,\mathrm{mrad}$ in the polarization of transmitted light and giant Kerr (reflection) rotations of $\approx \pi/2\,\mathrm{rad}$ in the polarization of very low frequency reflected light incident on a single slab \cite{slab1,slab2,slab3}. Furthermore, when the wavelength of the incident light is tuned to a reflection minimum (i.e. $kd = N\pi$ where $N$ is an integer) the Faraday rotation is universally quantized in units of the fine structure constant \cite{slab1,slab2}. This quantization highlights the difference between topological insulators and other magneto-electric materials such as chiral, Tellegen or gyrotropic media. In chiral media the wave equation includes a term that is proportional to the chirality and hence even in a homogeneous chiral material Faraday rotations will occur in the bulk, rather then just at the surface as in a topological insulator. Furthermore, chiral media are reciprocal whereas topological insulators are non-reciprocal. Tellegen media are non-reciprocal and the wave equation also has a term proportional to the gradient of the magneto-electric parameter but this parameter also contributes to a change in the bulk propagation, a change that is absent in topological insulators. Gyrotropic media are also non-reciprocal but are necessarily anisotropic as well. The displacement field is related to the electric field via the constitutive equation $\mathbf{D} = \varepsilon\mathbf{E} + i\mathbf{E}\times\mathbf{g}$ where $\mathbf{g}$ is the gyration vector, which, in turn, is related to a applied quasi-static magnetic field. Thus, there is no direct coupling between the electric and magnetic fields of a propagating electro-magnetic wave. This leads to the presence of Faraday rotation in the bulk rather then just on the surface. In addition, Faraday rotations in gyrotropic media are strongly dependent on frequency and hence do not show the universality that appears in topological insulators. Lastly, none of the media types listed above show the quantization in units of the fine-structure constant that topological insulators exhibit.

Until now experimental demonstration of this effect has been absent, in part owing to the small nature of the effect. However, recent work in $\mathrm{THz}$ spectroscopy of thin films subject to low temperatures and high magnetic fields have observed polarization rotations of $\approx 7\,\mathrm{mrad}$ consistent with the expected universally quantized value of $\alpha$ \cite{TIexp1,TIexp2,TIexp3}. For example, experiments in strained $\mathrm{HgTe}$ \cite{TIexp1} observed Faraday rotation of $\alpha$ for magnetic fields of $4\,\mathrm{T}$ and effective carrier temperature of $25\,\mathrm{K}$. The experiment also resolved the next highest plateau, indicated by a Faraday rotation of $\approx 3\alpha$, at a magnetic fields of $3\,\mathrm{T}$. Similar results have also been obtain for $\mathrm{Bi_{2}Se_{3}}$ thin films \cite{TIexp2} and $\mathrm{Cr}$ doped $\mathrm{(BiSb)_{2}Te_{3}}$ \cite{TIexp3}.

Following the first demonstrations of this effect, there will be an interest to study this effect in detail and enhancement of the small signal will make this task easier. Furthermore, this effect could prove to be useful in polarization control of $\mathrm{THz}$ radiation. However, a $\approx 7\,\mathrm{mrad}$ rotation is too small for practical applications. Thus, it is worth looking for systems that enhance this effect. As this effect only occur at interfaces, planar multi-layered structures (Bragg gratings) are a natural choice of system to explore.

\section{Transfer Matrix}

\begin{figure}[b]
\centering
\includegraphics[width=1\linewidth]{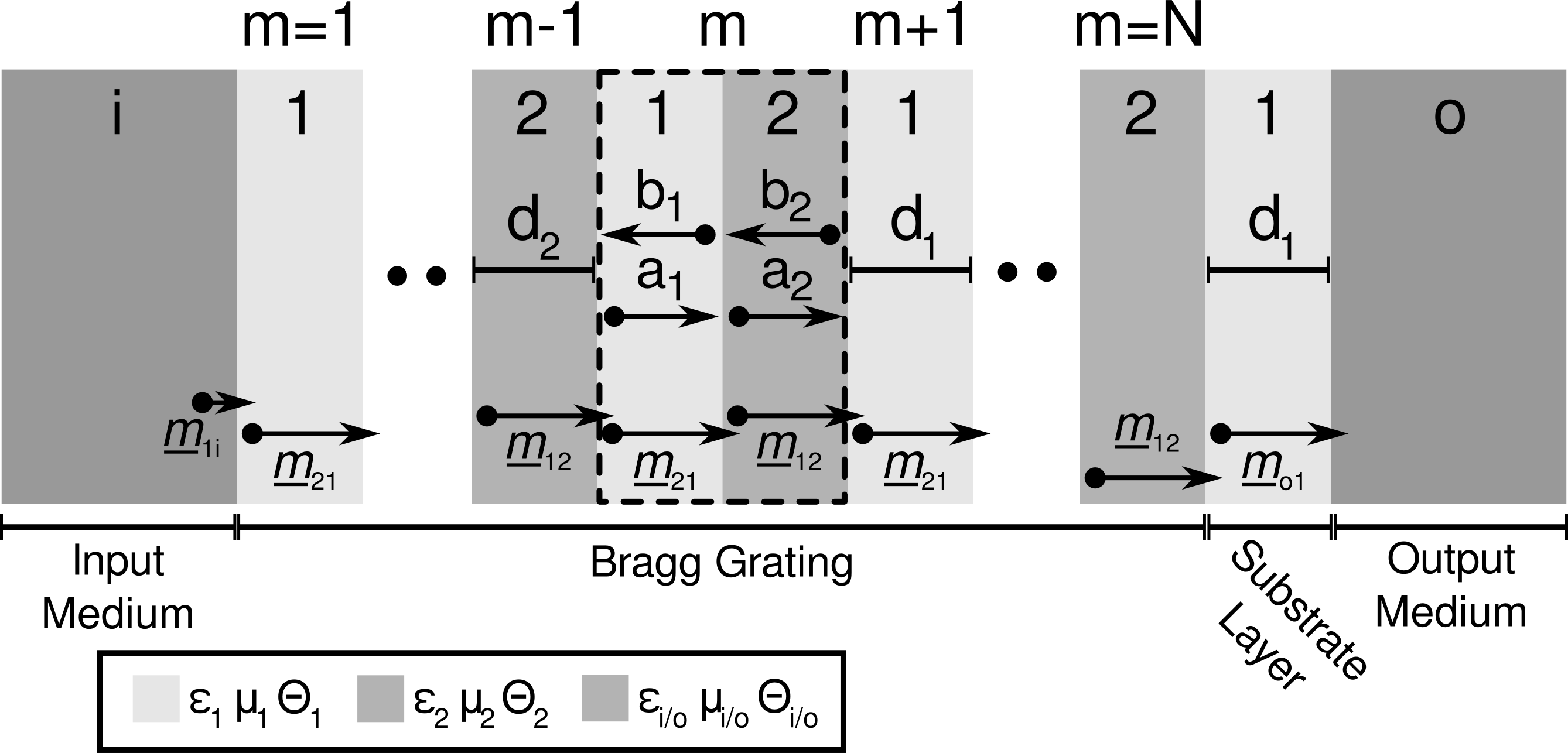}
\caption{A Bragg grating constructed from a layered axionic material and coupled to an input and output medium. The $\bm{\underline{m}}_{ij}$ matrices compute the right moving, $a_{i}$, and left moving, $b_{i}$, field amplitudes at the left hand interface in a layer with medium $i$ given the right moving, $a_{j}$, and left moving, $b_{j}$, field amplitudes at the left hand interface in the previous layer with medium $j$.}
\label{bragg}
\end{figure} 
Consider a Bragg grating, periodic in the $x$ direction, with layer thicknesses $d_{1}$ and $d_{2}$ ($d=d_{1}+d_{2}$) and corresponding permittivities $\varepsilon_{1}$ and $\varepsilon_{2}$, permeabilities $\mu_{1}$ and $\mu_{2}$ and axion couplings $\Theta_{1}$ and $\Theta_{2}$, respectively (See Fig \ref{bragg}). Light is incident from the left from an input medium with permittivity $\varepsilon_{i}$, permeability $\mu_{i}$ and axion couplings $\Theta_{i}$ and exits to the right first into a substrate layer with permittivity $\varepsilon_{1}$, permeability $\mu_{1}$ and axion couplings $\Theta_{1}$ (this ensures that the grating is symmetric) and then to an output medium with permittivity $\varepsilon_{o}$, permeability $\mu_{o}$ and axion couplings $\Theta_{o}$. The electric field ansatz reads
\begin{equation}
\mathbf{E}_{m,\nu}(\mathbf{r},\omega) = \left(\begin{array}{c}
E_{m, \nu, x}(x,\omega)\\
E_{m, \nu, y}(x,\omega)\\
E_{m, \nu, z}(x,\omega)
\end{array}\right)e^{ik_{p}z},
\label{E}
\end{equation}
where, $m$ labels the unit cell and $\nu \in 1,2$ indicates the medium layer. Here, $k_{p}$ is the wavevector parallel to the interfaces. Substituting the expression in Eq. \eqref{E} into Eq. \eqref{EHelmholtz} and noting that within each homogeneous region $\bm{\nabla}\Theta = 0$, one finds that the $y$ and $z$ components of the electric field components obey the usual $1D$ wave equation
\begin{equation}
\frac{\partial^{2}}{\partial x^{2}}E_{l,m,\nu}(x,\omega) + k_{\nu,x}^{2}E_{l,m,\nu}(x,\omega) = 0,
\end{equation}
where $l\in y,z$ and $k_{\nu, x} = \sqrt{k_{\nu}^{2} - k_{p}^{2}} = k_{\nu}\cos\phi_{\nu}$ with $k_{\nu}^{2} = \omega^{2}\mu_{\nu}\varepsilon_{\nu}$ and $\phi_{\nu}$ the incidence angle of the light with the interface. These equations admit the usual plane-wave solutions
\begin{gather}
E_{y,m,\nu}(x,\omega) = a_{y,m,\nu}e^{ik_{\nu,x}x} + b_{y,m,\nu}e^{-ik_{\nu,x}x},\\
E_{z,m,\nu}(x,\omega) = a_{z,m,\nu}e^{ik_{\nu,x}x} + b_{z,m,\nu}e^{-ik_{\nu,x}x},
\end{gather}
with the associated magnetic fields given by Eq. \eqref{con2}
\begin{align}
H_{y,m,\nu}(x,\omega) &= -\frac{k_{\nu}^{2}}{\omega\mu_{\nu} k_{\nu,x}}\left[a_{z,m,\nu}e^{ik_{\nu,x}x} - b_{z,m,\nu}e^{-ik_{\nu,x}x}\right]\nonumber\\
&\qquad - \frac{\alpha}{\pi}\Theta_{\nu}\left[a_{y,m,\nu}e^{ik_{\nu,x}x} + b_{y,m,\nu}e^{-ik_{\nu,x}x}\right],\\
H_{z,m,\nu}(x,\omega) &= \frac{k_{\nu,x}}{\omega\mu_{\nu}}\left[a_{y,m,\nu}e^{ik_{\nu,x}x} - b_{y,m,\nu}e^{-ik_{\nu,x}x}\right]\nonumber\\
&\qquad - \frac{\alpha}{\pi}\Theta_{\nu}\left[a_{z,m,\nu}e^{ik_{\nu,x}x} + b_{z,m,\nu}e^{-ik_{\nu,x}x}\right].
\end{align}
Here, $a_{i,m,\nu}$ and $b_{i,m,\nu}$ are the amplitudes of the right and left travelling waves, respectively.

Considering fundamental electromagnetic boundary conditions at the interfaces at $x=(m-1)d+d_{1}$ and $x=(m-1)d+d_{1}+d_{2}=md$, the amplitudes of the field at the left interface of each layer, given the amplitude of the field at the left interface of the previous layer, are found to be (See Fig \ref{bragg})
{\allowdisplaybreaks
\begin{gather}
\left(\begin{array}{c}
a_{y,m,2}\\
a_{z,m,2}\\
b_{y,m,2}\\
b_{z,m,2}
\end{array}\right) = \bm{\underline{m}}_{21}
\left(\begin{array}{c}
a_{y,m,1}\\
a_{z,m,1}\\
b_{y,m,1}\\
b_{z,m,1}
\end{array}\right),\\
\left(\begin{array}{c}
a_{y,m+1,1}\\
a_{z,m+1,1}\\
b_{y,m+1,1}\\
b_{z,m+1,1}
\end{array}\right) = \bm{\underline{m}}_{12}
\left(\begin{array}{c}
a_{y,m,2}\\
a_{z,m,2}\\
b_{y,m,2}\\
b_{z,m,2}
\end{array}\right),
\end{gather}
}
where the transfer matrices $\bm{\underline{m}}_{ij}$ read
\begin{widetext}
\begin{equation}
\bm{\underline{m}}_{ij} = 
\frac{1}{2}\left(\begin{array}{cccc}
\left(1+\gamma_{ji}^{TE}\right)e^{ik_{j,x}d_{j}} & \frac{\Delta\Theta_{ij}}{k_{i,x}}e^{ik_{j,x}d_{j}} & \left(1-\gamma_{ji}^{TE}\right)e^{-ik_{j,x}d_{j}} & \frac{\Delta\Theta_{ij}}{k_{i,x}}e^{-ik_{j,x}d_{j}}\\
-\frac{\Delta\Theta_{ij}}{k_{i,x}}\left(\frac{k_{i,x}^{2}}{k_{i}^{2}}\right)e^{ik_{j,x}d_{j}} & \left(1+\gamma_{ij}^{TM}\right)e^{ik_{j,x}d_{j}} & -\frac{\Delta\Theta_{ij}}{k_{i,x}}\left(\frac{k_{i,x}^{2}}{k_{i}^{2}}\right)e^{-ik_{j,x}d_{j}} & \left(1-\gamma_{ij}^{TM}\right)e^{-ik_{j,x}d_{j}}\\
\left(1-\gamma_{ji}^{TE}\right)e^{ik_{j,x}d_{j}} & -\frac{\Delta\Theta_{ij}}{k_{i,x}}e^{ik_{j,x}d_{j}} & \left(1+\gamma_{ji}^{TE}\right)e^{-ik_{j,x}d_{j}} & -\frac{\Delta\Theta_{ij}}{k_{i,x}}e^{-ik_{j,x}d_{j}}\\
\frac{\Delta\Theta_{ij}}{k_{i,x}}\left(\frac{k_{i,x}^{2}}{k_{i}^{2}}\right)e^{ik_{j,x}d_{j}} & \left(1-\gamma_{ij}^{TM}\right)e^{ik_{j,x}d_{j}} & \frac{\Delta\Theta_{ij}}{k_{i,x}}\left(\frac{k_{i,x}^{2}}{k_{i}^{2}}\right)e^{-ik_{j,x}d_{j}} & \left(1+\gamma_{ij}^{TM}\right)e^{-ik_{j,x}d_{j}}\\
\end{array}\right),
\label{mij}
\end{equation}
\end{widetext}
with $\gamma_{ij}^{TE} = k_{i,x}/k_{j,x}$, $\gamma_{ij}^{TM} = \varepsilon_{j}k_{i,x}/\varepsilon_{i}k_{j,x}$ and $\Delta\Theta_{ij} = \omega\mu_{j}\alpha(\Theta_{i}-\Theta_{j})/\pi$. In addition to these matrices we have the transfer matrices at the input, $\bm{\underline{m}}_{1i}$, and output, $\bm{\underline{m}}_{o1}$, faces of the grating. The latter can be found directly from Eq. \eqref{mij} whereas the former requires one to set $d_{j}\rightarrow 0$. By concatenating these matrices, one can find the field in any layer of the multi-layered system. Thus for an $N$ unit cell system with a substrate layer, field amplitudes on the input side of the grating are related to those on the output side via
\begin{equation}
\left(\begin{array}{c}
a_{y,o}\\
a_{z,o}\\
b_{y,o}\\
b_{z,o}
\end{array}\right) = \bm{\underline{m}}_{T}
\left(\begin{array}{c}
a_{y,i}\\
a_{z,i}\\
b_{y,i}\\
b_{z,i}
\end{array}\right),
\label{mTi}
\end{equation}
with $\bm{\underline{m}}_{T} = \bm{\underline{m}}_{o1}\cdot\left(\bm{\underline{m}}_{12}\cdot\bm{\underline{m}}_{21}\right)^{N}\cdot\bm{\underline{m}}_{1i}$.

\begin{figure}[t]
\centering
\includegraphics[width=0.5\linewidth]{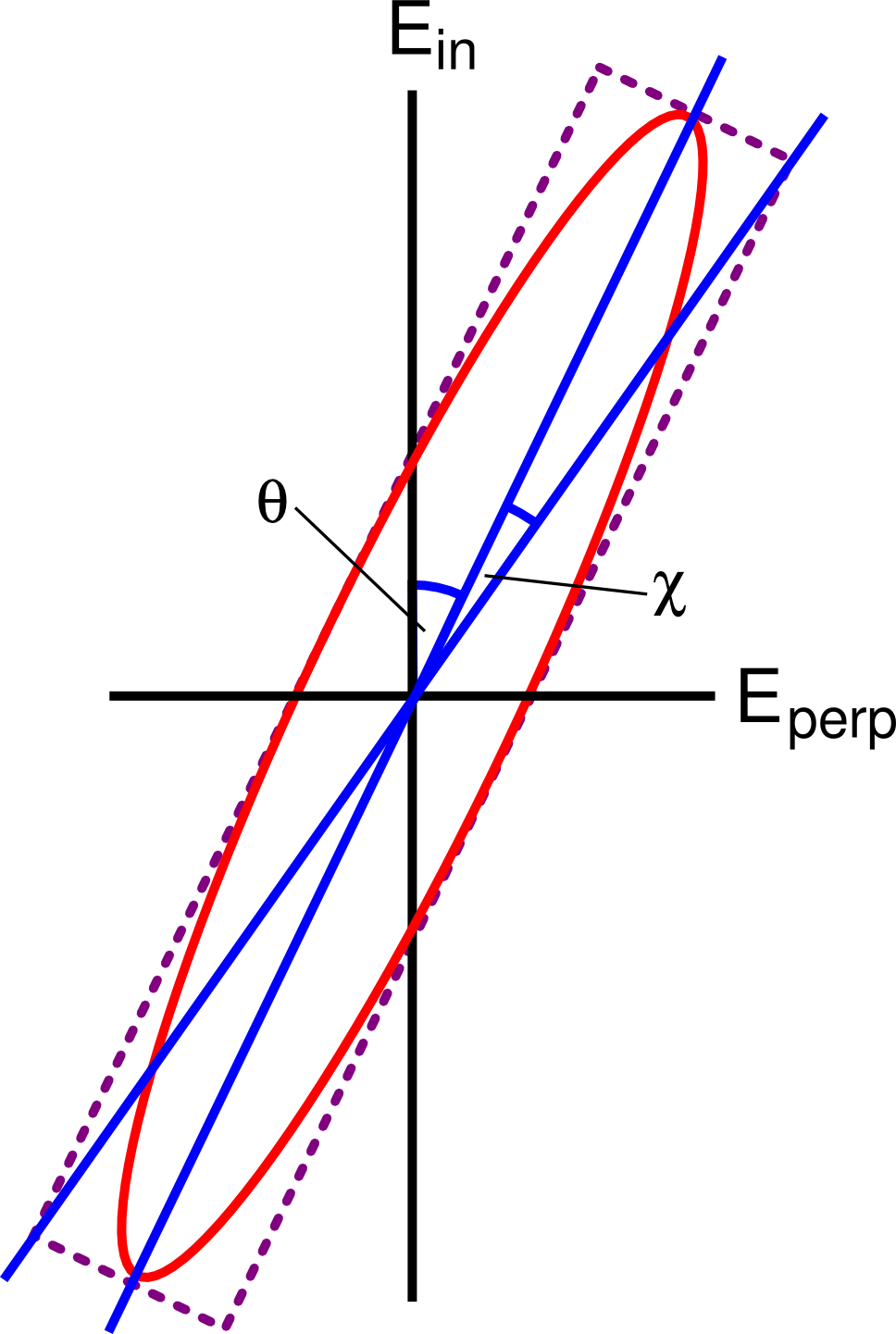}
\caption{(Color online) The rotation angle, $\theta$, and ellipticity, $\chi$, of elliptically polarized light.}
\label{Angles}
\end{figure}

Given input field amplitudes $a_{y,i}$ and $a_{z,i}$, the reflected and transmitted fields can be found found by setting $b_{y,o}=b_{z,o}=0$ (as there is no incident flux from the right) in Eq. \eqref{mTi} and solving for the remaining amplitudes.  The reflected power is given by
\begin{gather}
R = \frac{|b_{\mathrm{in},i}|^{2} + |b_{\mathrm{perp},i}|^{2}}{|a_{\mathrm{in},i}|^{2}},
\end{gather}
and the Faraday angles and associated ellipticity can be found from
\begin{gather}
\theta_{F} = \frac{1}{2}\arctan\left(\frac{2|a_{\mathrm{in},o}|^{2}}{|a_{\mathrm{in},o}|^{2}-|a_{\mathrm{perp},o}|^{2}}\mathrm{Re}\left[\frac{a_{\mathrm{perp},o}}{a_{\mathrm{in},o}}\right]\right),
\end{gather}
and
\begin{gather}
\chi_{F} = \frac{1}{2}\arcsin\left(\frac{2|a_{\mathrm{in},o}|^{2}}{|a_{\mathrm{in},o}|^{2}+|a_{\mathrm{perp},o}|^{2}}\mathrm{Im}\left[\frac{a_{\mathrm{perp},o}}{a_{\mathrm{in},o}}\right]\right),
\end{gather}
respectively, where $in\in y,z$ refers to the polarization of the input light and $perp\in y,z$ refers the polarization perpendicular to it (See Fig. \ref{Angles}).

As an example, we consider an $8$ unit cell ($16$-layer) Bragg grating consisting of alternating layers of $\mathrm{Si}$ and $\mathrm{Bi_{2}Se_{3}}$ with varying values of $d_{1}$ and $d_{2}$ ($d = d_{1} + d_{2} = 1$). Furthermore, we assume the light is incident from the vacuum and exits to a $\mathrm{Si}$ substrate layer and then to the vacuum. Thus, $\varepsilon_{i} = \varepsilon_{o} = 1$, $\varepsilon_{1} = 12$ and $\varepsilon_{2} = 16$. (The static permittivity of $\mathrm{Bi_{2}Se_{3}}$ is large but for wavelengths on the order of a few micron this lower value is appropriate \cite{bi2se3}). In the following we will assume that magnetic effects are small and hence $\mu_{1} = \mu_{2} = \mu_{i} = \mu_{o} = 1$. 

\begin{figure}[b]
\centering
\includegraphics[width=1\linewidth]{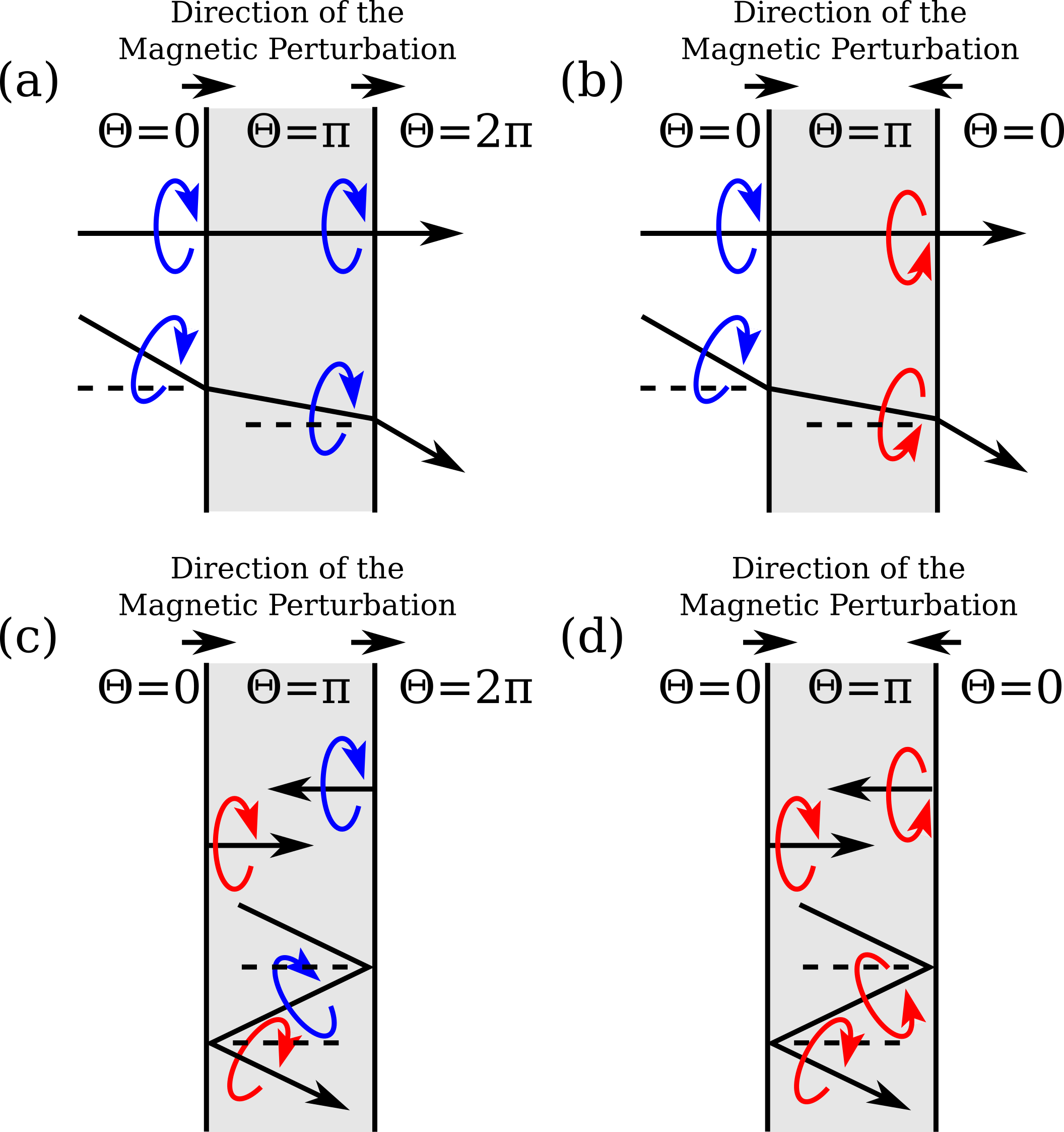}
\caption{(Color online) The orientation of the rotation of the light polarization on reflection and transmission at an interface. The red arrow indicates a clockwise rotation and the blue arrow a counter-clockwise rotation with respect to the propagation direction of the light. (a) The optical path for transmitted light at normal and oblique incidence angles in the parallel configuration. In this case the Faraday rotations are in the same direction. (b) The optical path for transmitted light at normal and oblique incidence angles in the anti-parallel configuration. In this case the Faraday rotations are in opposite direction. (c) The optical path for reflected light at normal and oblique incidence angles in the parallel configuration. In this case the Faraday rotations are in the same direction. (d) The optical path for reflected light at normal and oblique incidence angles in the anti-parallel configuration. In this case the Faraday rotations are in opposite directions.}
\label{P_AP}
\end{figure} 
There are two situations to consider. The first is the parallel configuration where the magnetic perturbation is in the same direction at each interface. This leads to an increase in the axion coupling in each successive layer. Hence, $\Theta_{i} = 0$, $\Theta_{1} = (2m-2)\pi$, $\Theta_{o} = 2N\pi$ and $\Theta_{2} = (2m-1)\pi$ [see Fig. \ref{P_AP} (a) and (c)]. The second is the anti-parallel configuration where the magnetic perturbation is in the opposite direction at each consecutive interface. This leads to the axion coupling alternating between values in each successive layer. Hence, $\Theta_{i} = \Theta_{1} = \Theta_{o} = 0$ and $\Theta_{2} = \pi$  [see Fig. \ref{P_AP} (b) and (d)]. 

\section{Results}

\subsection{Normal Incidences}

Before discussing high incidence angles it is worth briefly reviewing the Faraday rotation results for normal incidences. At normal incidences the $TE$ and $TM$ polarizations are indistinguishable. As the axion coupling is on the order of the fine structure constant, and hence $\approx 10^{-2}$, the change in reflectivity compared to a non-axionic Bragg grating is negligible.
\begin{figure*}[t]
\centering
\includegraphics[width=1\linewidth]{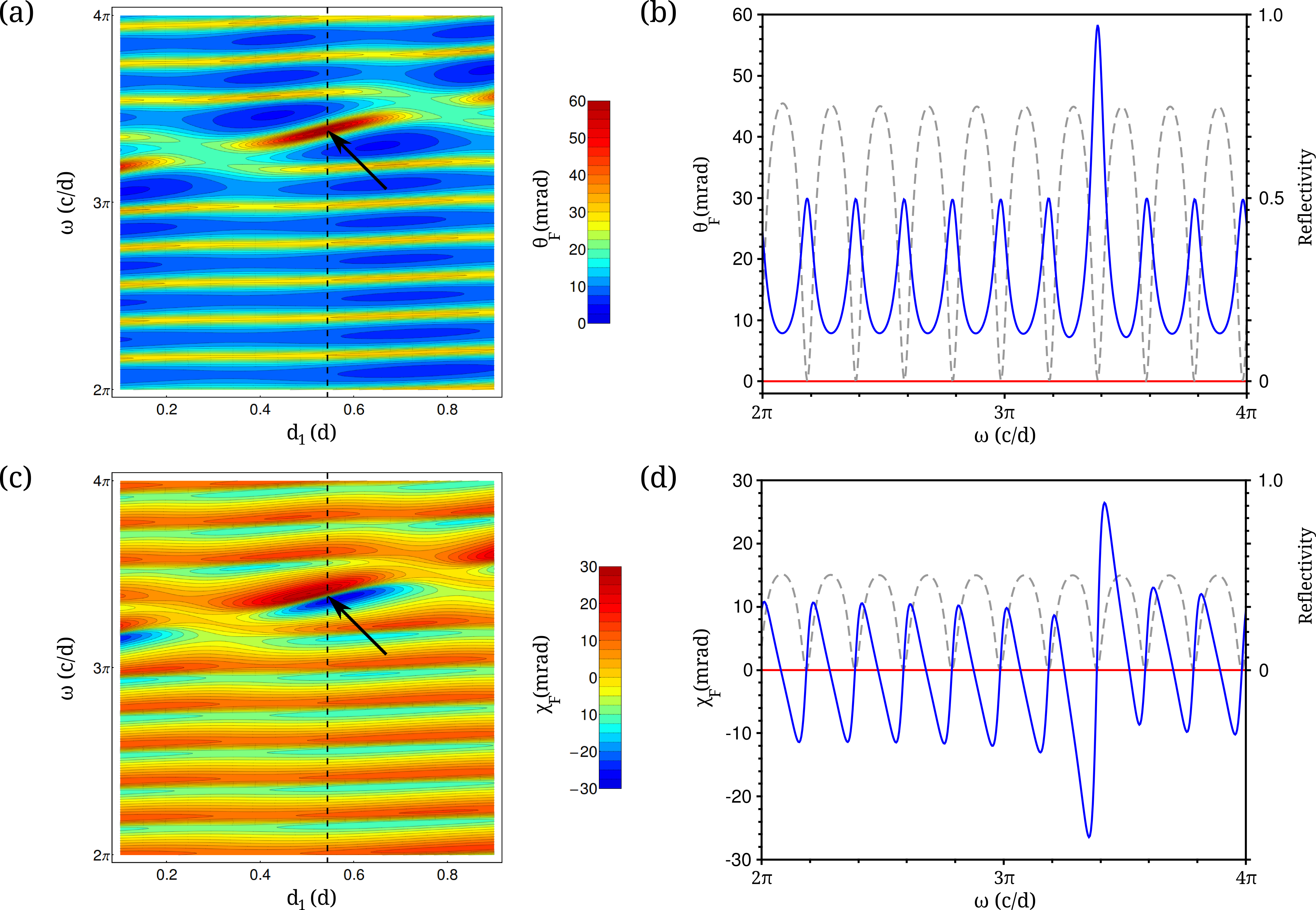}
\caption{(Color online) (a) The Faraday rotation as a function of frequency and grating structure at normal incidence angle for the parallel configuration. The arrow marks the maximum Faraday rotation of $\approx 8\alpha$. (b) The Faraday rotation as a function of frequency at normal incidence angle for the parallel (blue) and anti-parallel (red) configuration for $d_{1} = \sqrt{\varepsilon_{2}}/(\sqrt{\varepsilon_{1}}+\sqrt{\varepsilon_{2}})$, marked by the black dashed line in (a). The gray dashed line indicates the reflectivity of the grating. (c) The ellipticity as a function of frequency and grating structure at normal incidence angle for the parallel configuration. The arrow marks the point of vanishing ellipticity coincident with the location of the maximum Faraday rotation. (d) The ellipticity as a function of frequency at normal incidence angle for the parallel (blue) and anti-parallel (red) configuration for $d_{1} = \sqrt{\varepsilon_{2}}/(\sqrt{\varepsilon_{1}}+\sqrt{\varepsilon_{2}})$, marked by a black dashed line in (c).}
\label{norm}
\end{figure*}
In the parallel configuration the change in the axion coupling at each interfaces is always $\pi$ and, hence, the Faraday rotation is always in the same direction [See Fig \ref{P_AP} (a)]. Furthermore, reflected waves will undergo the same rotation as a transmitted wave [See Fig \ref{P_AP} (c)]. Hence, both reflected left and right propagating waves will be rotated in the same direction as the transmitted right propagating wave. Thus the rotations at each interface, whether in reflection or transmission, sum and one should observe a non-zero Faraday rotation. In the anti-parallel configuration the change in the axion coupling at each interfaces alternates between $\pi$ and $-\pi$. Transmission at alternate interface causes a finite rotation in opposite directions [See Fig \ref{P_AP} (b)]. As before, reflected waves are rotated in the same direction as the transmitted waves [See Fig \ref{P_AP} (d)]. Thus the rotations at each interface, whether in reflection or transmission, cancel and one should not observe Faraday rotation for normal incidence angles. 

Figure \ref{norm} (a) shows the Faraday rotation for the parallel configuration as a function of frequency and grating structure. The maximum rotation of $58\,\mathrm{mrad}$, or equivalently $\approx 8\alpha$, occurs when the grating layers are $d_{1} = \sqrt{\varepsilon_{2}}/(\sqrt{\varepsilon_{1}}+\sqrt{\varepsilon_{2}})$ and $d_{2} = \sqrt{\varepsilon_{1}}/(\sqrt{\varepsilon_{1}}+\sqrt{\varepsilon_{2}})$ and frequency $2\pi/(d_{1}\sqrt{\varepsilon_{1}} + d_{1}\sqrt{\varepsilon_{1}})$. For this grating structure and frequency the amplitude of the wave at the interfaces of the grating is maximized and, hence, the rotation is maximized. Furthermore, we see that, for these parameters, the grating leads to a Faraday rotation of $\alpha$ per unit cell and, hence, we are in the quantized regime. This shows that this regime, not only leads to a universally quantized rotation, but also that this rotation is maximal. Figure \ref{norm} (b) shows the Faraday rotation for the parallel and anti-parallel configurations. For the parallel configuration one sees a non-zero Faraday rotation at all frequencies whereas the Faraday rotation for the anti-parallel configuration vanishes at all frequencies. The suppression of the Faraday rotation in the region of high reflectivity is due to the destructive interference reducing the amplitude of the wave in the region of the interfaces, i.e. the location at which the rotation occurs. Figure \ref{norm} (c) shows the ellipticity for the parallel configuration as a function of frequency and grating structure. The arrow marks the ellipticity at the point of maximum Faraday rotation. One can see that at this point the ellipticity vanishes. Figure \ref{norm} (d) shows the ellipticity for the parallel and anti-parallel configurations. A similar phenomenology to the Faraday rotation is observed with a non-zero ellipticity for the parallel configuration and a vanishing ellipticity for the anti-parallel configuration. The increased ellipticity in the region of high reflectivity is due to the multiple scattering inside the grating causing a greater phase shift compared to regions of low reflectivity.
 
\subsection{Oblique Incidences}

We now consider high incidence angles. At non-normal incidence angles the $TE$ and $TM$ polarizations are distinguishable. The first aspect to note is that, owing to Snell's law, the incidence angle and the angle of transmission are not equal. As the axionic interaction is a function of the in-plane fields [c.f. Eq \eqref{EHelmholtz}], Faraday rotation will have a different magnitude at alternate interfaces. Thus, in the anti-parallel configuration, there will not be perfect cancellation and, hence, one should see a non-vanishing Faraday rotation. The second aspect to note is that the axion interaction increases if the in-plane magnetic field is increased with respect to the in-plane electric field \cite{TIgreen}. As the incidence angle is increased, the in-plane magnetic field of the $TE$ mode is reduced while the in-plane electric field remains constant whereas for the $TM$ mode, the in-plane electric field is reduced while the in-plane magnetic field remains constant. Thus, the $TM$ mode should show strong enhancement of the Faraday rotation whereas the $TE$ mode should display suppression.

\begin{figure}[b]
\centering
\includegraphics[width=0.97\linewidth]{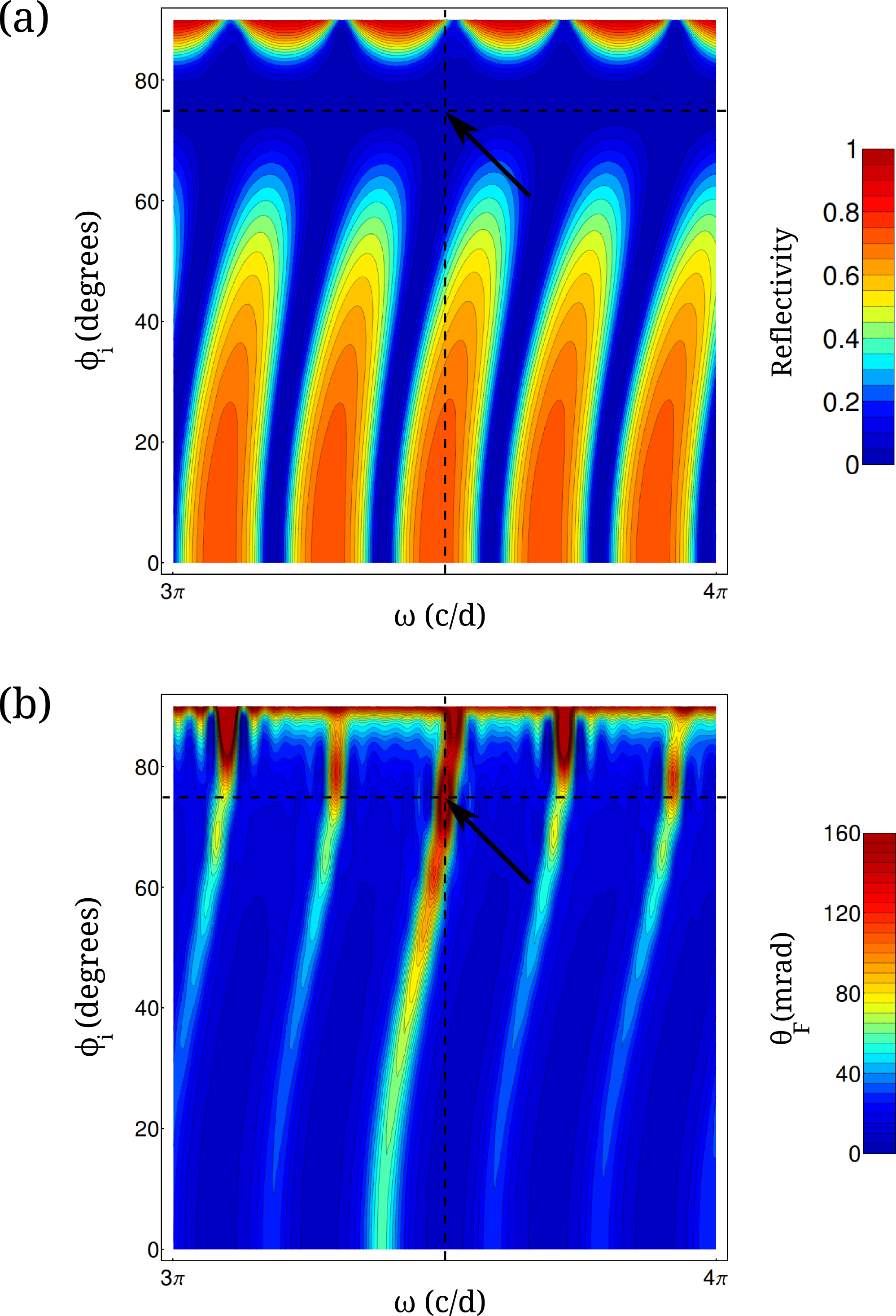}
\caption{(Color online) (a) The reflectivity of the grating as a function of frequency and incidence angle for the $TM$ polarization in the parallel configuration. The horizontal dashed line marks $76^{\mathrm{o}}$ in the high transmission window near the Brewster angle. The vertical dashed line marks the frequency $1.754\,c/d$ which corresponds to the maximum Faraday rotation for this incidence angle. This point is marked by the arrow. (b) The Faraday rotation as a function of frequency and incidence angle for the $TM$ polarization in the parallel configuration. The horizontal dashed line marks $76^{\mathrm{o}}$ and vertical dashed line marks the frequency $1.754\,c/d$. The maximum Faraday rotation is marked by the arrow. In this cases $d_{1} = \sqrt{\varepsilon_{2}}/(\sqrt{\varepsilon_{1}}+\sqrt{\varepsilon_{2}})$ and $d_{2} = \sqrt{\varepsilon_{1}}/(\sqrt{\varepsilon_{1}}+\sqrt{\varepsilon_{2}})$.}
\label{con}
\end{figure} 
Figure \ref{con} (a) shows the reflectivity for the grating as a function of incidence angle and frequency. As before $d_{1} = \sqrt{\varepsilon_{2}}/(\sqrt{\varepsilon_{1}}+\sqrt{\varepsilon_{2}})$ and $d_{2} = \sqrt{\varepsilon_{1}}/(\sqrt{\varepsilon_{1}}+\sqrt{\varepsilon_{2}})$. One can see a marked decrease in the reflectivity as one approaches the Brewster angle. In this region one can expect high transmission. Figure \ref{con} (b) shows the Faraday rotation as a function of incidence angle and frequency. Within the region of high transmission one can see a maximum in the Faraday rotation. This occurs at a frequency of $1.754\,c/d$ and an incidence angle of $76^{\mathrm{o}}$. The Faraday rotation at this point is found to be $\approx 232\,\mathrm{mrad}$ ($\approx 13^{\mathrm{o}}$) and the reflectivity is $5.7\%$. Such a rotation is larger than has been seen in gyrotropic gratings (e.g. \cite{gyro}) and comparable to the rotation in chiral gratings (e.g. \cite{chiral}).

\begin{figure*}[t]
\centering
\includegraphics[width=0.97\linewidth]{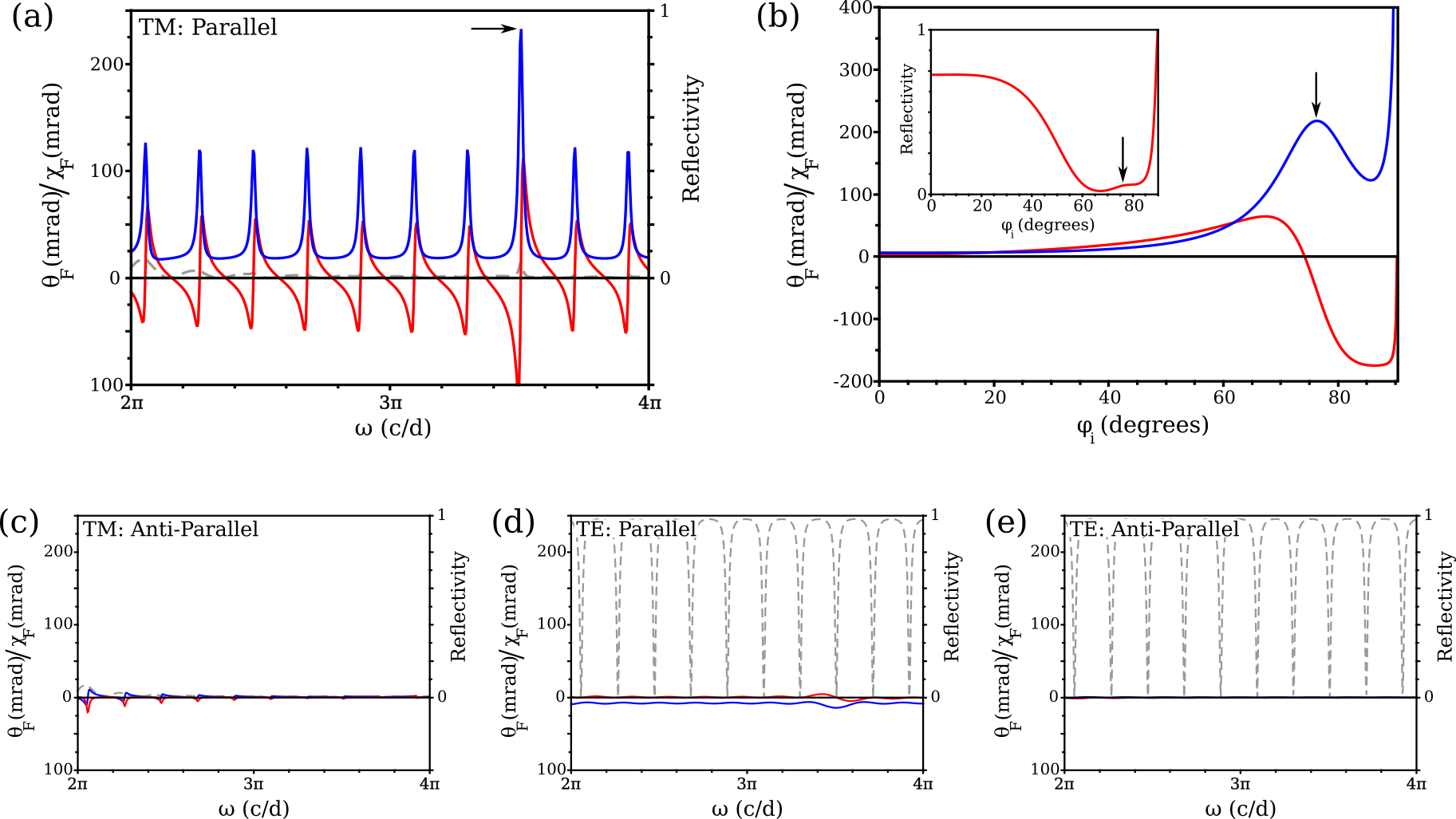}
\caption{(Color online) (a) The Faraday rotation (blue) and ellipticity (red) of as a function of frequency for the $TM$ polarization in the parallel configuration for a $76^{\mathrm{o}}$ incidence angle. The gray dashed line indicates the reflectivity of the grating. The arrow marks the frequency $1.754\,c/d$. (b) The Faraday rotation (blue) and ellipticity (red) of as a function of incidence angle for the $TM$ polarization in the parallel configuration with frequency $1.754\,c/d$. The arrow indicates the local maximum in the Faraday rotation at $76^{\mathrm{o}}$. (inset) The reflectivity of the grating as a function of incidence angle. The arrow marks the transmission window at $76^{\mathrm{o}}$ incidence angle. (c), (d), (e) The Faraday rotation (blue) and ellipticity (red) of as a function of frequency for the $TM$ polarization anti-parallel and $TE$ polarization parallel and anti-parallel configurations, respectively, for a $76^{\mathrm{o}}$ incidence angle. The gray dashed lines indicate the reflectivity of the grating. In all cases $d_{1} = \sqrt{\varepsilon_{2}}/(\sqrt{\varepsilon_{1}}+\sqrt{\varepsilon_{2}})$ and $d_{2} = \sqrt{\varepsilon_{1}}/(\sqrt{\varepsilon_{1}}+\sqrt{\varepsilon_{2}})$.}
\label{angle}
\end{figure*} 
Figure \ref{angle} (a), (c)-(e) show the Faraday rotation and ellipticity for the $TM$ and $TE$ polarizations in the, parallel and anti-parallel configurations, respectively, for a $76^{\mathrm{o}}$ incidence angle and $d_{1} = \sqrt{\varepsilon_{2}}/(\sqrt{\varepsilon_{1}}+\sqrt{\varepsilon_{2}})$ and $d_{2} = \sqrt{\varepsilon_{1}}/(\sqrt{\varepsilon_{1}}+\sqrt{\varepsilon_{2}})$. One sees the large Faraday rotation of $\approx 232\,\mathrm{mrad}$ and a small ellipticity of $\approx 32\,\mathrm{mrad}$ for the $TM$ polarization for light of a frequency $1.754\,c/d$ while the Faraday rotation of the $TE$ polarization is between $\approx 1-10\,\mathrm{mrad}$. This is an enhancement of a factor of $4$ in the Faraday rotation compared to the $TM$ polarization normal incidences and orders of magnitude higher than the $TE$ polarization at any incidence angle. Note that, for normal incidences, as one is in the quantized regime, adding additional unit cells would increase the Faraday rotation by a factor of $\alpha \approx 7\,\mathrm{mrad}$ per cell. Hence, the increase obtained from the higher incidence angle is much larger than could be obtained by just adding more layers. Comparison of Figure \ref{angle} (a) with Figure \ref{norm} (b) shows that the enhancement is uniform across all frequencies and, thus, any effects that are present in the spectrum at normal incidence angles will be present at oblique angles. Hence, the universal quantization of the Faraday rotation will still be present though the quanta of rotation would now be larger than $\alpha$. However, one would still be able to observe plateaus in the Faraday rotation (which correspond to different hall plateaus or, equivalently, different axion couplings) as one varies the applied magnetic field or magnetic surface impurity density.

Figure \ref{angle} (b) shows that the $76^{\mathrm{o}}$ incidence angle is a local maximum in the Faraday rotation. The enhancement of the parallel configuration over the anti-parallel configuration is, again, due to the additive nature of the rotations in the former structure as compared to the partial cancellation of the rotations in the latter. With such a large rotation, these structures offer an ideal system in which to observe the magneto-electric effect by optical means. The inset in Fig. \ref{angle} (b) shows that the local maximum in Faraday rotation at frequency $1.754\,c/d$ lies in a region of high transmission, with a reflectivity at the $76^{\mathrm{o}}$ incidence angle of $5.7\%$. At even larger incidence angles the Faraday rotation and ellipticity are even higher, however, the increasing reflectivity at such high incidence angles means that the signal from these large rotations will become increasingly hard to detect. Despite this, Bragg gratings offer a sizeable enhancement of the Faraday rotation and provide a structure where the magneto-electric effect can be more easily detected.

Finally it is worth noting that this enhancement in the Faraday rotation in the $TM$ polarization in the parallel configuration over the $TE$ polarization and anti-parallel configurations is independent of frequency and hence is not a spectral separation process such as has been studied in chiral slabs where birefringent mirrors lead to a Fano resonance that separates positive and negative spin modes of photons \cite{Fano}. Furthermore, As the effect is related to the ratio of the in-plane magnetic field electric field ratio \cite{TIgreen} the opposite effect, enhancement of the $TE$ mode over the $TM$ mode, is not possible. Hence, within topological insulators there is a fundamental asymmetry between the $TM$ and $TE$ modes. This in itself could prove to be a useful effect for polarization control in future electro-magentic devices.

\section{Summary}

Here we have shown, using a transfer matrix approach, that the magneto-electric effect induced Faraday rotation of plane polarized light incident on a time-symmetry-broken topological insulator can be enhanced by a factor of $4$ by using a $TM$ polarized light at high incidence angles. The enhancement is uniform across all frequencies and hence any features of the spectrum at normal incidences will still be observable at oblique incidences. Thus, features such as the quantized Hall plateaus will be, not only observable but, in fact, enhanced. The inherent asymmetry in the enhancement, favouring $TM$ polarizations over $TE$ polarizations could prove useful in novel electro-magnetic devices where distinguishability the two modes is important.

\section{Acknowledgements}

This work is supported by the Singapore National Research Foundation under NRF Grant No. NRF-NRFF2011-07.



\begin{thebibliography}{99}

\bibitem{TIrev1}
M. Z. Hasan and C. L. Kane, Rev. Mod. Phys. \textbf{82}, 3045 (2010).

\bibitem{TIrev2}
X.-L. Qi and S.-C. Zhang, Rev. Mod. Phys. \textbf{83}, 1057 (2011).

\bibitem{3DTI1}
L. Fu and C. L. Kane, Phys. Rev. B \textbf{76}, 045302 (2007).

\bibitem{3DTI2}
D. Hsieh, D. Qian, L. Wray, Y. Xia, Y. S. Hor, R. J. Cava and M. Z. Hasan, Nature \textbf{452}, 970 (2008).

\bibitem{3DTI3}
H. Zhang, C.-X. Liu, X.-L. Qi, X. Dai, Z. Fang and S.-C. Zhang, Nat. Phys. \textbf{5}, 438 (2009).

\bibitem{3DTI4}
C.-X. Liu, X.-L. Qi, H. Zhang, X. Dai, Z. Fang and S.-C. Zhang, Phys. Rev. B \textbf{82}, 045122 (2010).

\bibitem{monopole}
X.-L. Qi, R. Li, J. Zang and S.-C. Zhang, Science \textbf{323}, 1184 (2009).

\bibitem{TITFT}
X.-L. Qi, T. L. Hughes and S.-C. Zhang, Phys. Rev. B \textbf{78}, 195424 (2008).

\bibitem{obukhov}
Y. N. Obukhov and F. W. Hehl, Phys. Lett. A \textbf{341}, 357 (2005).

\bibitem{wilczek}
F. Wilczek, Phys. Rev. Lett \textbf{58}, 1799 (1987).

\bibitem{cr}
J. Zhang, C.-Z. Chang, P. Tang, Z. Zhang, X. Feng, K. Li, L.-L. Wang, X. Chen, C. Liu, W. Duan, K. He, Q.-K. Xue, X. Ma and Y. Wang, Science \textbf{339}, 1582 (2013).

\bibitem{qi}
J. Maciejko, X.-L. Qi, H. D. Drew and S.-C. Zhang, Phys. Rev. Lett. \textbf{105}, 166803 (2010).

\bibitem{chang}
M.-C. Chang and M.-F. Yang, Phys. Rev. B \textbf{80}, 113304 (2009).

\bibitem{slab1}
W.-K. Tse and A. H. MacDonald, Phys. Rev. B \textbf{82}, 161104(R) (2010).

\bibitem{slab2}
W.-K. Tse and A. H. MacDonald, Phys. Rev. B \textbf{84}, 205327 (2011).

\bibitem{slab3}
W.-K. Tse and A. H. MacDonald, Phys. Rev. Lett. \textbf{105}, 057401 (2010).

\bibitem{TIexp1}
V. Dziom, A. Shuvaev, A. Pimenov, G. V. Astakhov, C. Ames, K. Bendias, J. B\"{o}ttcher, G. Tkachov, E. M. Hankiewicz, C. Br\"{u}ne, H. Buhmann and L. W. Molenkamp, arXiv:1603.05482 [cond-mat.str-el].

\bibitem{TIexp2}
L. Wu, M. Salehi, N. Koirala, J. Moon, S. Oh and N. P. Armitage, arXiv:1603.04317 [cond-mat.mes-hall].

\bibitem{TIexp3}
K. N. Okada, Y. Takahashi, M. Mogi, R. Yoshimi, A. Tsukazaki, K. S. Takahashi, N. Ogawa, Masashi Kawasaki and Y. Tokura, arXiv:1603.02113 [cond-mat.mes-hall].

\bibitem{bi2se3}
O. Madelung, U. R\"{o}ssler, M. Schulz, \textit{SpringerMaterials: Bismuth selenide (Bi2Se3) optical properties, dielectric constants}, (Springer-Verlag, Heidelberg, 1998).

\bibitem{TIgreen}
J. A. Crosse, S. Fuchs and S. Y. Buhmann, Phys. Rev. A \textbf{92}, 063831 (2015).

\bibitem{gyro}
Y. H. Lu, M. H. Cho, J. B. Kim, G. J. Lee, Y. P. Lee and J. Y. Rhee, Opt. Express \textbf{16}, 5378 (2008).

\bibitem{chiral}
A. Potts, W. Zhang and D. M. Bagnall, Phys. Rev. A \textbf{77}, 043816 (2008).

\bibitem{Fano}
X. Piao, S. Yu, J. Hong and N. Park, Sci. Rep. \textbf{5}, 16585 (2015).

\end{thebibliography}
\end{document}